# Incommensurately modulated twin structure of nyerereite $Na_{1.64}K_{0.36}Ca(CO_3)_2$

Nadezhda B. Bolotina[1*], Pavel N. Gavryushkin[2,3], Andrey V. Korsakov[2,3], Sergey V. Rashchenko[2,3], Yurii V. Seryotkin[2,3], Alexander V. Golovin[2,3], Bertrand N. Moine[4], Anatoly N. Zaitsev[5,6], Konstantin D. Litasov[2,3]

[1] Shubnikov Institute of Crystallography of Federal Scientific Research Centre "Crystallography and Photonics" of Russian Academy of Sciences, Leninskii prosp. 59, 119333 Moscow, Russia
[2] Sobolev Institute of Geology and Mineralogy, Siberian Branch of Russian Academy of Sciences, prosp. acad. Koptyuga 3, 630090 Novosibirsk, Russia
[3] Novosibirsk State University, Pirogova st. 2, 630090 Novosibirsk, Russia
[4] Laboratoire Magmas et Volcans UMR6524 CNRS Université J. Monnet, 23 rue du Dr. P. Michelon F-42023 Saint-Etienne, France
[5] Department of Mineralogy, St. Petersburg State University, University Emb. 7/9, 199034 St. Petersburg, Russia
[6] Department of Earth Sciences, Natural History Museum, Cromwell Road, London, SW7 5BD, UK

\* **Contact author**:

Dr. Nadezhda B. Bolotina

Shubnikov Institute of Crystallography of Federal Scientific Research Centre "Crystallography and Photonics" of Russian Academy of Sciences, Leninskii prosp. 59, 119333 Moscow, Russia
Email: bolotina@ns.crys.ras.ru

**Abstract**

Incommensurately modulated twin structure of nyerereite $Na_{1.64}K_{0.36}Ca(CO_3)_2$ has been first determined in the (3+1)D symmetry group $Cmcm(\alpha00)00s$ with modulation vector $\mathbf{q} = 0.383\mathbf{a}^*$. Unit-cell values are $a = 5.062(1)$, $b = 8.790(1)$, $c = 12.744(1)$ Å. Three orthorhombic components are related by threefold rotation about [001]. Discontinuous crenel functions are used to describe occupation modulation of Ca and some $CO_3$ groups. Strong displacive modulation of the oxygen atoms in vertexes of such $CO_3$ groups is described using x-harmonics in crenel intervals. The Na, K atoms occupy mixed sites whose occupation modulation is described by two ways using either complementary harmonic functions or crenels. The nyerereite structure has been compared both with commensurately modulated structure of K-free $Na_2Ca(CO_3)_2$ and with widely known incommensurately modulated structure of $\gamma$-$Na_2CO_3$.

**Introduction**

Nyerereite, a carbonate mineral of approximate chemical composition $(Na,K)_2Ca(CO_3)_2$ was discovered in the alkali-rich carbonate lavas erupted by the Oldoinyo Lengai volcano, Tanzania (Dawson, 1962a, 1962b). First structure study of this mineral was performed by McKie & Frankis (1977) but some principal structural features remained unknown until now because of incommensurate structure modulation accompanied with twinning. The modulation phenomenon in $(Na,K)_2Ca(CO_3)_2$ attracts special attention because of its composition resemblance to γ-$Na_2CO_3$, which is well-known as the very first incommensurately modulated structure that was solved using a superspace approach (Wolff, 1974; Aalst, Holander & Peterse, 1976). Incommensurate structure of γ-$Na_2CO_3$ was re-refined later more than once (Dusek *et al.*, 2003; Arakcheeva & Chapuis, 2005 and references therein).

McKie & Frankis (1977) studied the room-temperature orthorhombic structure of nyerereite $Na_{1.64}K_{0.36}Ca(CO_3)_2$ in the *Cmc*$2_1$ group ($a = 5.044$, $b = 8.809$, $c = 12.743$ Å, $Z = 4$) whereas its high-temperature modification was determined as hexagonal one in the *P*$6_3$*mc* group ($a = 5.05$, $c = 12.85$ Å, $Z = 2$). Non-Bragg reflections *hkl* with $h = n \pm 0.383m$ ($m = 1$, rarely 2) but *k*, *l* integers were found on precession photographs of the room-temperature phase that indicated incommensurate structural modulation along the *a*-axis with the wave vector **q** = 0.383**a**\*. The superspace approach was very new in 1977 and the average structure of nyerereite was described in the conventional 3D model without taking satellites into account. Positions of some oxygen atoms were not determined and positions of cations were determined only approximately as a result. In present work, the incommensurately modulated crystal structure of the 3-component twin crystal of nyerereite is first determined in a (3+1)D model.

**Experimental**

Oldoinyo Lengai natrocarbonatite active volcano in northern Tanzania is situated along the western flank of the East African Rift Valley. At present, it is the only active volcano in the world erupting extrusive carbonate lava. Oldoinyo Lengai is a young dominantly phobolitic-nephelinitic volcano, which is less than 790,000 years old (Sherrod *et al.*, 2013), formed by a complex sequence of events, including explosive eruptions of nephelinitic and phonolitic ashes and effusion of mixed assemblages of silicate and carbonate lavas. In the recent past, several major Plinian eruptions have occurred, e.g. 1960-1966, 2007-2008 (Mitchell & Dawson, 2007); between these eruptions the activity has remained relatively constant, essentially effusive and fumarolic, with minor carbonatite lava flows within the northern crater. Crater activity between 1993 and 1997 has been well summarized by Nyamweru (1997). The sample for this study was collected during eruption of October 1995 within the active hornito, named T37. Since that time the sample was kept in the vacuum chamber to prevent the destruction of the sample. Similarly to other natrocarbonatite lavas from Oldoinyo Lengai the sample consists of phenocrysts of nyerereite and gregoryite in a finer-grained carbonate-sylvite-fluorite matrix (Fig.1).

Chemical composition of nyerereite was obtained by wavelength dispersive X-ray spectroscopy (WDS) using JEOL JXA-8100 electron microprobe analysis at the Sobolev Institute of Geology and Mineralogy (Novosibirsk, Russia). The results of chemical analysis are summarized in Table 1. The average mineral formula according to performed analysis is the same as one determined by McKie & Frankis (1977): $Na_{1.64}K_{0.36}Ca(CO_3)_2$. As appears from 61 analyses, potassium content is almost constant both within one grain and between different grains varying in the range 0.36–0.37 a.p.f.u. Compared with published data the studied mineral is similar in composition with nyerereite from 1992 and 1993 lava eruptions (Zaitsev *et al.*, (2009).

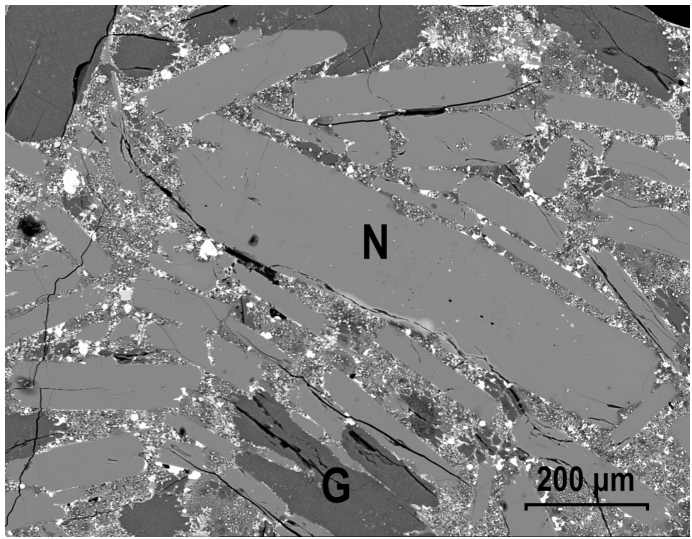

**Figure 1**. Backscattered electron image of natrocarbonatite sample. Light-grey prismatic phenocrysts – nyerereite (one of the grains marked N), dark-grey rounded phenocrysts - gregoryite (one of the grains marked G).

**Table 1.** Representative chemical analyses (WDS) of nyerereite from Oldoinyo Lengai natrocarbonatite. Each analysis is from individual nyerereite grain.

| Wt % | 1 | 2 | 3 | 4 | 5 | 6 | 7 | 8 | 9 | 10 |
|---|---|---|---|---|---|---|---|---|---|---|
| $Na_2O$ | 23,35 | 24,15 | 23,70 | 23,45 | 23,49 | 23,86 | 23,84 | 24,09 | 23,88 | 24,12 |
| $K_2O$ | 8,00 | 7,84 | 7,68 | 8,05 | 7,78 | 7,80 | 7,93 | 7,82 | 7,74 | 7,72 |
| CaO | 23,92 | 24,33 | 23,96 | 24,28 | 24,42 | 24,31 | 24,31 | 24,06 | 23,93 | 24,23 |
| SrO | 2,25 | 1,69 | 2,28 | 1,85 | 1,87 | 2,12 | 1,90 | 2,27 | 2,13 | 1,82 |
| BaO | 1,10 | 0,63 | 0,72 | 0,67 | 0,74 | 0,55 | 0,56 | 0,68 | 0,68 | 0,57 |
| MgO | 0,09 | 0,00 | 0,04 | 0,03 | 0,02 | 0,01 | 0,00 | 0,00 | 0,01 | 0,03 |
| FeO | 0,03 | 0,04 | 0,03 | 0,00 | 0,02 | 0,03 | 0,04 | 0,02 | 0,02 | 0,04 |
| MnO | 0,12 | 0,10 | 0,12 | 0,10 | 0,12 | 0,15 | 0,13 | 0,08 | 0,07 | 0,13 |
| $SO_3$ | 1,06 | 1,02 | 1,01 | 1,02 | 1,03 | 1,48 | 0,96 | 1,10 | 1,17 | 0,99 |
| $P_2O_5$ | 0,32 | 0,19 | 0,29 | 0,29 | 0,49 | 0,37 | 0,47 | 0,18 | 0,21 | 0,15 |
| Cl | 0,30 | 0,28 | 0,27 | 0,34 | 0,23 | 0,30 | 0,30 | 0,30 | 0,29 | 0,34 |
| -O=F,Cl | 0,07 | 0,06 | 0,06 | 0,08 | 0,05 | 0,07 | 0,07 | 0,07 | 0,07 | 0,08 |
| Total | 60,46 | 60,20 | 60,03 | 60,00 | 60,16 | 60,91 | 60,37 | 60,53 | 60,06 | 60,06 |
| Structural formulae based on cations positive charge +4 | | | | | | | | | | |
| Na | 1,635 | 1,678 | 1,661 | 1,643 | 1,644 | 1,659 | 1,660 | 1,675 | 1,674 | 1,678 |
| K | 0,369 | 0,358 | 0,354 | 0,371 | 0,358 | 0,357 | 0,363 | 0,358 | 0,357 | 0,353 |
| Ca | 0,926 | 0,934 | 0,928 | 0,940 | 0,944 | 0,934 | 0,935 | 0,924 | 0,927 | 0,931 |
| Sr | 0,047 | 0,035 | 0,048 | 0,039 | 0,039 | 0,044 | 0,040 | 0,047 | 0,045 | 0,038 |
| Ba | 0,016 | 0,009 | 0,010 | 0,009 | 0,010 | 0,008 | 0,008 | 0,010 | 0,010 | 0,008 |
| Mg | 0,005 | 0,000 | 0,002 | 0,002 | 0,001 | 0,001 | 0,000 | 0,000 | 0,000 | 0,002 |
| Fe | 0,001 | 0,001 | 0,001 | 0,000 | 0,001 | 0,001 | 0,001 | 0,000 | 0,000 | 0,001 |
| Mn | 0,004 | 0,003 | 0,004 | 0,003 | 0,004 | 0,004 | 0,004 | 0,002 | 0,002 | 0,004 |
| Total | 3,002 | 3,018 | 3,008 | 3,007 | 3,001 | 3,008 | 3,012 | 3,016 | 3,016 | 3,016 |
| S | 0,029 | 0,028 | 0,027 | 0,028 | 0,028 | 0,040 | 0,026 | 0,030 | 0,032 | 0,027 |
| P | 0,010 | 0,006 | 0,009 | 0,009 | 0,015 | 0,011 | 0,014 | 0,006 | 0,007 | 0,005 |
| Cl | 0,018 | 0,017 | 0,017 | 0,021 | 0,014 | 0,018 | 0,018 | 0,018 | 0,018 | 0,020 |
| Total | 0,057 | 0,050 | 0,053 | 0,058 | 0,057 | 0,069 | 0,058 | 0,053 | 0,056 | 0,052 |
| $CO_3$ | 1,95 | 1,96 | 1,95 | 1,95 | 1,94 | 1,93 | 1,94 | 1,95 | 1,95 | 1,96 |
| Total | 2,004 | 2,006 | 2,004 | 2,006 | 2,000 | 2,003 | 2,002 | 2,006 | 2,006 | 2,008 |
| $CO_2$ wt% | 39,48 | 39,98 | 39,52 | 39,49 | 39,42 | 39,51 | 39,64 | 39,90 | 39,49 | 39,93 |
| Total | 99,94 | 100,18 | 99,55 | 99,50 | 99,58 | 100,42 | 100,01 | 100,43 | 99,55 | 99,99 |

Small piece of the sample was crushed down to 100 microns and only optically clean fragments of crystals were selected for further X-ray study. A 0.05×0.10×0.14 mm³ crystal for X-ray diffraction experiment was selected using a polarizing microscope. Data collection was performed at Xcalibur Gemini Ultra diffractometer (Oxford Diffraction) with Enhance (Mo) source and Ruby CCD-detector. CrysAlisPro software, Agilent Technologies, Version 1.171.37.35h (release 09-02-2015 CrysAlis171.NET) was used for the data reduction. An orthorhombic basic lattice with the cell parameters $a$ = 5.062(1), $b$ = 8.790(1), $c$ = 12.744(1) Å was determined. Modulation vector was refined using the *nada qvector* command of CrysAlisPro as **q** = 0.383(1)**a**\* in full agreement with McKie & Frankis (1977). Corresponding first-order satellites *hklm* were observed in the (*hkl*) planes (*l* = 1, 2…) of the 3D diffraction pattern in three directions at the |**q**| length from the main *hkl*0 reflections. The studied crystal was assumed therefore to be a twin whose three components were related by threefold rotation about [001] axis. Data reduction from incommensurate crystal was performed three times separately for each twin component. Three data sets were uploaded to the Jana2006 program (Petříček, Dušek & Palatinus, 2014) for the structure determination and scaled in one scale using the main reflections only. Data has not been averaged because of twinning. The structure has been determined in the (3+1)D symmetry group $Cmcm(\alpha 00)00s$, the modulation vector **q** = 0.383(1)**a**\*. Main features of X-ray diffraction experiment and structure refinement are summarized in Table 2.

**Table 2.** Experimental and structure refinement details.
___________________________________________________

Crystal data

| | |
|---|---|
| Chemical formula | $Na_{1.64}K_{0.36}Ca(CO_3)_2$ |
| $M_r$ | 211.9 |
| Crystal system, space group | Orthorhombic, $Cmcm(\alpha 00)00s$ † |
| Temperature (K) | 293 |
| Modulation wavevector | **q** = 0.383(1)**a**\* |
| $a$, $b$, $c$ (Å) | 5.062(1), 8.790(1), 12.744(1) |
| $V$ (Å$^3$) | 567.0(1) |
| $Z$ | 4 |
| Radiation type | X-ray, $\lambda$ - 0.7107 Å |
| $\mu$ (mm$^{-1}$) | 1.465 |
| Crystal size (mm) | 0.05×0.10×0.14 |

Data collection

| | |
|---|---|
| Diffractometer | Xcalibur Ruby Gemini Ultra |
| Absorption correction | multi-scan |
| $T_{min}$, $T_{max}$ | 0.823, 1.000 |
| No. of reflections measured | 37213 |
| $(\sin \theta/\lambda)_{max}$ (Å$^{-1}$) | 0.745 |

Refinement details

| | |
|---|---|
| Refinement technique | LSQ based on $F$ |
| Weighting sceme | $w = 1/[\sigma^2(F_{obs}) + (0.01F_{obs})^2]$ |
| No. of all reflections (main + satellites) | 37213 (5599 + 31614) |
| No. of observed [$I > 3\sigma(I)$] reflections (main + satellites) | 4400 (2261+2139) |
| $R_{obs}$ (all, main, satellites), % | 6.99, 5.82, 9.47 |
| $wR_{obs}$ (all, main, satellites), % | 7.49, 5.74, 12.62 |
| $GOF_{obs}$ | 1.39 |
| No. of parameters | 79 |
| No. of restraints | 5 |
| No. of constraints | 13 |
| $\Delta\rho_{max}$, $\Delta\rho_{min}$ (e Å$^{-3}$) | 4.23. -4.76 |

† Symmetry codes: (i) $x1$; $x2$; $x3$; $x4$; (ii) $-x1$; $-x2$; $x3+0.5$; $-x4+0.5$; (iii) $-x1$; $x2$; $-x3+0.5$; $-x4$; (iv) $x1$; $-x2$; $-x3$; $x4+0.5$; (v) $-x1$; $-x2$; $-x3$; $-x4$; (vi) $x1$; $x2$; $-x3+0.5$; $x4+0.5$; (vii) $x1$; $-x2$; $x3+0.5$; $x4$; (viii) $-x1$; $x2$; $x3$; $-x4+0.5$ plus (0.5; 0.5; 0; 0) translation.

**Discussion**

Potassium-free crystals of Na$_2$Ca(CO$_3$)$_2$ were synthesized recently by hydrothermal method (Gavryushkin *et al.*, 2016), and their crystal structure was determined in the orthorhombic space group $P2_1ca$; $a$ = 10.0713(5), $b$ = 8.7220(2), $c$ = 12.2460(4) Å, $Z$ = 8. A $5a \times c$ fragment of hydrothermal Na$_2$Ca(CO$_3$)$_2$ is shown in Fig. 2.

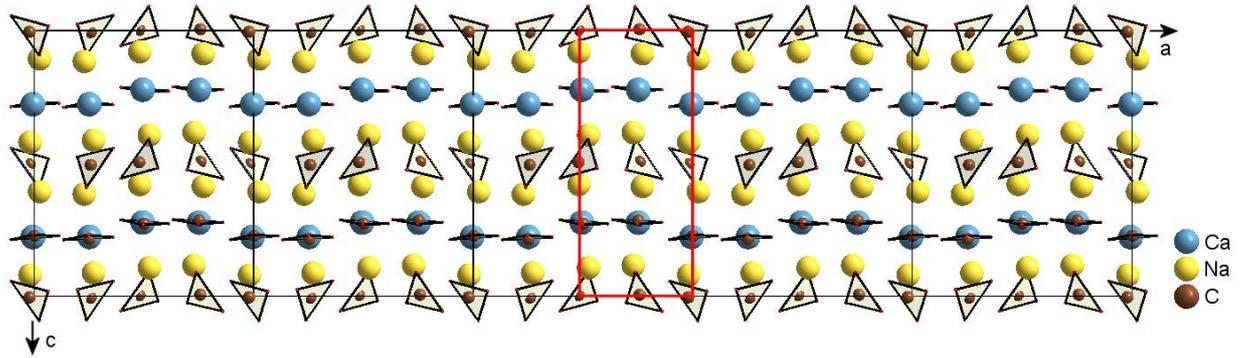

**Figure 2**. A fragment (five unit cells) of synthetic Na$_2$Ca(CO$_3$)$_2$ in the projection on (010). Contours of smaller cell suitable for a (3+1)D structure description are shown by red colour.

Two types of atomic layers alternate along the *c*-axis. Layers of the first type are built from Ca atoms and CO$_3$-groups parallel to the (*a*, *b*) plane, which figure as horizontal CO$_3$ groups hereafter. Each layer of the second type is a sandwich formed by Na atoms and vertical CO$_3$ groups between them. The motif of structure modulation along the *a*-axis is quite clear. Ca atoms and horizontal carbonate groups modulate in-phase changing their *z*-coordinates abruptly. Sodium atoms undergo smoother *z*-modulation accompanied with a noticeable *x*-modulation. The vertical CO$_3$ groups modulate in most complicated manner. Obviously, the structure of the synthetic crystal could be described as commensurately modulated one using the ($a_c$=$a$/2, $b_c$=$b$, $c_c$=$c$) unit cell and wave vector **q** = 0.5**a**$_c$*.

One may expect similar motif of positional modulation from the nyerereite structure keeping in mind, however, that its modulation vector is $\mathbf{q} = 0.383\mathbf{a}^* \cong 2/5\mathbf{a}^*$. It means that two modulation periods extend for about five unit cells along the *a*-axis but we failed to arrange the vertical $CO_3$ groups even roughly in the 3D space just moving into the (5*a*, *b*, *c*) supercell. The problem has been resolved only after conversion to a superspace model in the following way.

The (3+1)D symmetry of the incommensurate nyerereite structure was described using the orthorhombic unit cell ($a = 5.062(1)$, $b = 8.790(1)$, $c = 12.744(1)$ Å) and the centrosymmetric (3+1)D symmetry group *Cmcm*($\alpha$00)00*s*. A possibility of using the centrosymmetric *Pbca* group instead of *P2$_1$ca* for hydrothermal $Na_2Ca(CO_3)_2$ crystals was discussed at length in (Gavryushkin *et al.*, 2016). Main argument in favour of non-centrosymmetric group was there that out of about 26000 reflections with I > 3$\sigma$, nearly 400 reflections (~1.5%) broke the reflection conditions of *Pbca* space group, while only 5 reflections close to 3$\sigma$ broke the reflection conditions of *P2$_1$ca* space group. So rigorous criteria, being applied to the defect nyerereite structure, would lead directly to a primitive monoclinic group like *P*11*m* owing to 125 main and satellite reflections with I > 3$\sigma$ breaking both the *C*-centering and the reflection condition $h0l : l = 2N$. So radical symmetry decreasing seems to be a poor idea concerning this crystal. One should notice, however, that two other samples of nyerereite, whose diffraction patterns we were able just to see but not to treat being no owners of them, revealed more significant distortions of the C-centering. As for the sample under study, it was analyzed first in the non-centrosymmetric (3+1)D symmetry group *C2cm*($\alpha$00)*s*0*s*. A strong probability of centrosymmetric group became evident after an origin shift by the amount of 0.25 was performed along the *x*1 axis, so that additional shift of the origin by the amount of -0.25 along the *x*4 axis provided the standard setting for *Cmcm*($\alpha$00)00*s*. The symmetry operators are listed under Table 2.

One Ca site, one mixed Na/K site plus six sites, which are occupied by members of one horizontal (C1, O1, O2, O2') and one vertical (C2, O3, O4, O4') carbonate group, form the symmetry-independent set in the *Cmcm*($\alpha$00)00*s* group of symmetry. The difference between this set and similar one in the *C2cm*($\alpha$00)*s*0*s* group is that in the non-centrosymmetric group all six vertexes of two carbonate groups are independent, whereas in the centrosymmetric group the third vertex O2' of the horizontal $CO_3$ group is coupled with O2 by a symmetry operator as well as O4' is coupled with O4 in the vertical $CO_3$ group.

Continuous displacive modulations of some of these atoms, namely C2, O1, O2, can be well described by harmonic functions determined on the whole period $0 \leq x4 \leq 1$. Other atomic sites, namely Ca, C1, O3, O4 undergo discontinuous modulations. It is strange at first glance that in the horizontal carbonate group the C1 position is modulated discontinuously whereas positional modulations of O1, O2 are defined by harmonics but it is so, indeed. The C1 atom changes its z-coordinate unevenly following Ca whereas z-coordinates of O1, O2 stick around 0.25 and need no discontinuous function to be properly described. In the vertical carbonate group, on the contrary, positional modulation of C2 is quite weak and may be well described by harmonics whereas the modulations of O3 and particularly O4 are strong, complicate and discontinuous.

These sites were assumed to be occupied with 100% and 0% probabilities respectively on one and other half of modulation period. It was described using crenel functions $o(x40, \Delta)$ with invariant $\Delta = 0.5$ for all these atoms. Variables $x40$ and $\Delta$ are assigned hereafter to the centre and the length of the crenel interval, respectively. The symmetry group $Cmcm(\alpha00)00s$ contains the operator ($x1$, $x2$, $-x3+0.5$, $x4+0.5$). It means that all the atoms, whose sites are determined by aforesaid crenels, undergo discontinuous $x3 \rightarrow -x3+0.5$ changes at $x4 = x40 \pm 0.25$. Actually, the original $C2cm(\alpha00)s0s$ group was build 'by hand'. After the unity operator ($x1$, $x2$, $x3$, $x4$), its first non-trivial operator ($x1$, $x2$, $-x3+0.5$, $x4+0.5$) was constructed simply looking at commensurately modulated structure of hydrothermal $Na_2Ca(CO_3)_2$. Another (3+1)D operator ($x1$, $-x2$, $x3+1/2$, $x4$) was based on the ($x$, $-y$, $z+0.5$) operator of the $Na_2Ca(CO_3)_2$ symmetry. Third non-trivial operator ($x1$, $-x2$, $-x3+1/2$, $x4+0.5$) was added to complete the set. These operators plus the (0.5, 0.5, 0, 0) translation formed $C2cm(\alpha00)s0s$ group whose further transformation to $Cmcm(\alpha00)00s$ was described above. The $Cmcm(\alpha00)00s$ group symbol is defined for the standard $Amam(00\gamma)s00$ group No 63.1.13.6 (Stokes, Campbell & van Smaalen, 2011; see also http://stokes.byu.edu/iso/ssg.php) after its transformation to the $\overline{c}ba$ setting. The non-standard setting is chosen to keep the sequence of the cell parameters like that in (McKie & Frankis, 1977). All the discontinuous changes happen simultaneously in a real 3D crystal, whose structure is the $t = 0$ section of the (3+1)D model. Since $x4 = t + \mathbf{q}\cdot\mathbf{r}(x1, x2, x3) + 0.25$, the $x40$ variables have been 'synchronized' using the equation $x40 = 0.383*x1 + 0.25$. Most of the $x1$ coordinates are zeros when the nyerereite structure is described in $Cmcm(\alpha00)00s$, so these $x40$ values are set to 0.25 for all the atoms except O4.

The Ca atoms are only ones among the atoms with discontinuous occupation modulation that do not undergo appreciable displacive modulations in crenel intervals (Fig. 3).

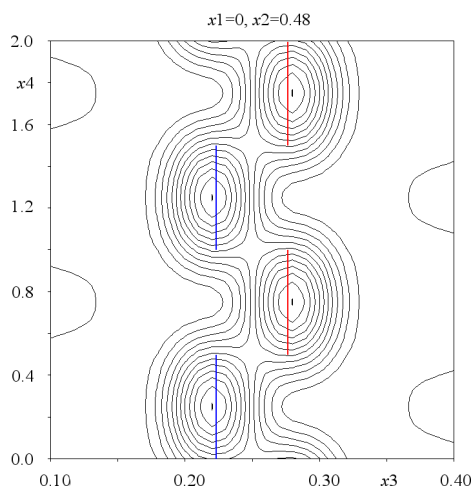

**Figure 3**. Observed Fourier map of the electron-density distribution in the neighborhood of the Ca site in the layer ($x3$, $x4$). Symmetry-related Ca($x1$, $x2$, $x3$, $x4$) and Ca($x1$, $x2$, $-x3+0.5$, $x4+0.5$) are differently coloured. Contour intervals are 5 e Å$^{-3}$.

Displacive modulations of C1, O3, O4 were described using x-harmonics in crenel intervals (Petříček *et al*., 2016). Two x-harmonics were used for C1 and four ones for O3, O4. Besides, the C–O and O–O distances in vertical $CO_3$ groups had to be restrained at the appropriate values of 1.28 and 2.22 Å, respectively, using the *distfix* command of Jana2006. Strong $x2$-modulations of O4 ($x1$, $x2$, $x3$, $x4$) and O4 ($x1$; $-x2$; $-x3$; $x4+0.5$) (b) over two modulation periods are shown in Fig. 4a, b. The O4 site (Fig. 4a), whose average coordinates are (~0.14, ~0.48, 0.04), may be split into O41(~0.17, ~0.56, ~0.05) and O42(~0.11, ~0.40, 0.03). Crenel interval $\Delta = 0.5$ is divided by two as a result, and separate modulation functions are determined in the adjacent crenel intervals $\Delta 1 = 0.25$ and $\Delta 2 = 0.25$, whose centers are synchronized at ~0.17*0.383 + 0.375 and ~0.11+0.383 + 0.125 for O41 and O42, respectively. Displacive modulation in each of two crenel intervals is described using two members in the x-harmonic expansion instead of four x-harmonics used to describe the modulation of O4. Such split model seems to be more appropriate as compared with initial one. Symmetry-related atoms O4 ($x1$, $x2$, $x3$, $x4$) and O4'($-x1$; $x2$; $x3$; $-x4+0.5$) form a side of an $O_3$ triangle but some O4–O4' distances are too short, so forced distance restrictions have to be applied whereas the O41–O42' and O41'–O42 distances are longer. It should be noticed, however, that the *distfix* command works well and crystal structure of nyerereite looks on much the same being refined both in the O4-split and O4-non-split models. The *R*-values in the non-split model for all, main and satellite reflections are $R_{obs} = 7.10$, 5.90, 9.66% and $wR_{obs} = 7.57$, 5.87, 12.63, i.e. just a little higher as compared with those presented in Table 2 for the split model, moreover that 73 parameters are refined in the non-split model but not 79 as in the split one.

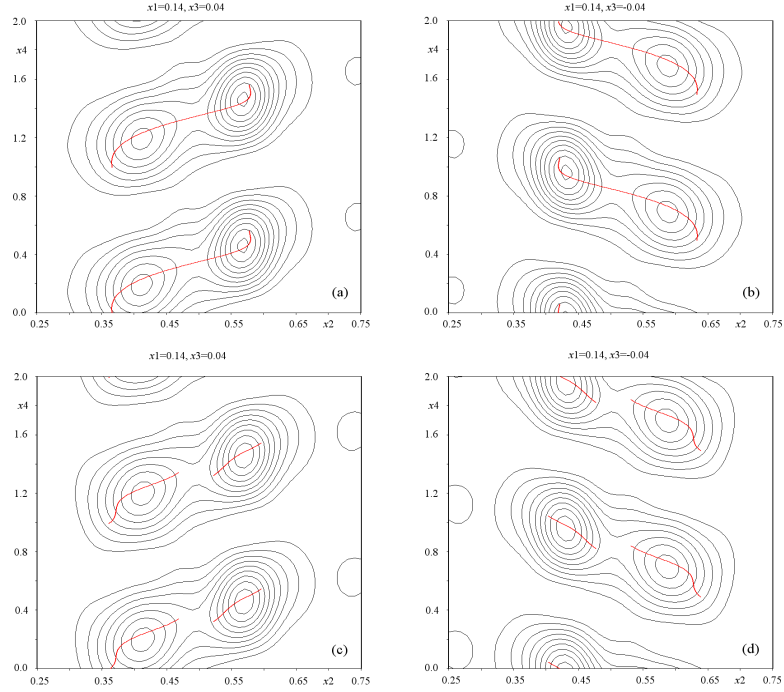

**Figure 4**. Observed ($x2$, $x4$) Fourier maps of the electron-density distribution in the neighborhood of O4 ($x1$, $x2$, $x3$, $x4$) (a) and O4 ($x1$; -$x2$; -$x3$; $x4$+0.5) in the O4-non-split (a, b) and O4-split (c, d) models. Each plot is obtained by $x1$-summation over the range $0.11 \leq x1 \leq 0.17$ and $x3$-summation over the ranges $0.03 \leq x3 \leq 0.05$ and $-0.05 \leq x3 \leq -0.03$ for the left and right maps, respectively, in increments of 0.01. Contour intervals are 0.5 e A$^{-3}$.

Displacive modulation of the mixed (Na, K) site can be well characterized by one harmonic determined on the whole $x4$-period as shown in Fig. 5a. Anticipated (Na, K) occupation modulation has been taken into consideration in two different ways. Model 1 assumes smooth variation of the potassium content at the mixed (Na, K) site. Complementary occupation modulation is described in one-harmonic approximation (Fig. 5b). In alternative model 2, the occupation modulation is described by crenels with Δ=0.82 for Na and Δ=0.18 for K. Centres of the crenel intervals are fixed at $x4_{0Na} = 0.25$ and at $x4_{0K} = x4_{0Na} + 0.5 = 0.75$ as it was determined from Fig. 5b.

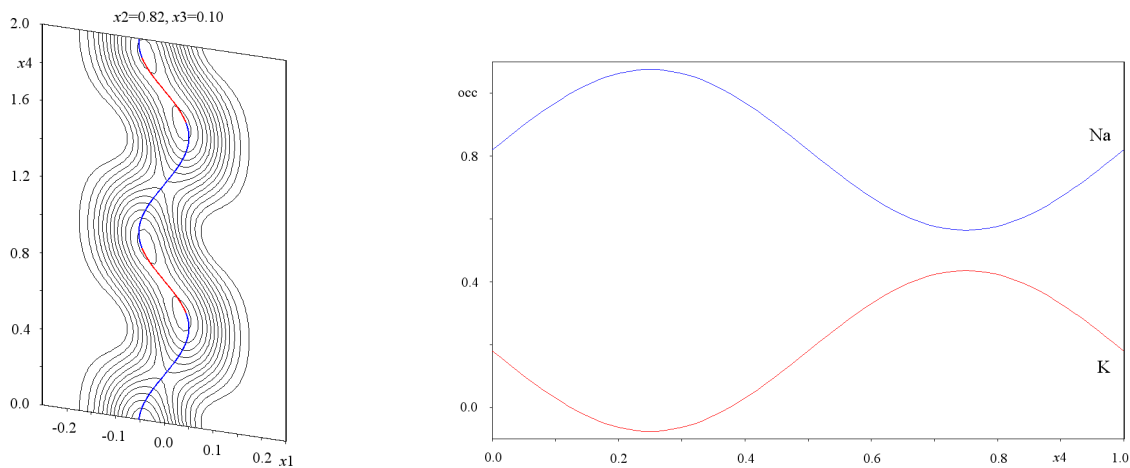

**Figure 5.** (a) Observed Fourier map of the electron-density distribution in the neighborhood of mixed (Na,K) sites in the layers ($x1$, $x4$). Segments occupied by K with probabilities more than 0.3 are coloured by red. Contour intervals are 1 e A$^{-3}$. (b) Complementary Na/K occupation modulation.

Values of $R_{obs}$ (all, main, satellites) = 6.99, 5.80, 9.50% as well as $wR_{obs}$ (all, main, satellites) = 7.58, 5.77, 12.86% obtained in model 2 are almost equal with those obtained in model 1 (see Table 2), so $R$-values themselves cannot be a criterion for the choice unicity. On the one hand, strong Na–K separation seems to be somewhat artificial, since there are no fundamental grounds to eliminate the possibility that Na and K can switch sometimes. On the other hand, an argument may be given in favour of the separation that ionic radius of K (1.33 Å) differs essentially from that of Na (0.98 Å). It is clear at any rate that just the Na–K distribution on the modulation period determines incommensurate character of structural modulation in whole. Besides, the Na–K distribution has a crucial influence on the modulation of oxygen atoms at vertexes of the $CO_3$ groups.

The twin component ratio was refined as 0.675(1) : 0.074(1) : 0.251(1). It differs from the results obtained for hydrothermal $Na_2Ca(CO_3)_2$, whose twin components are nearly equal in volume. The 3D approximants of the nyerereite structure obtained in two models are represented in Fig. 6(a, b). Variable content of K is shown in Fig. 6a by variable colours.

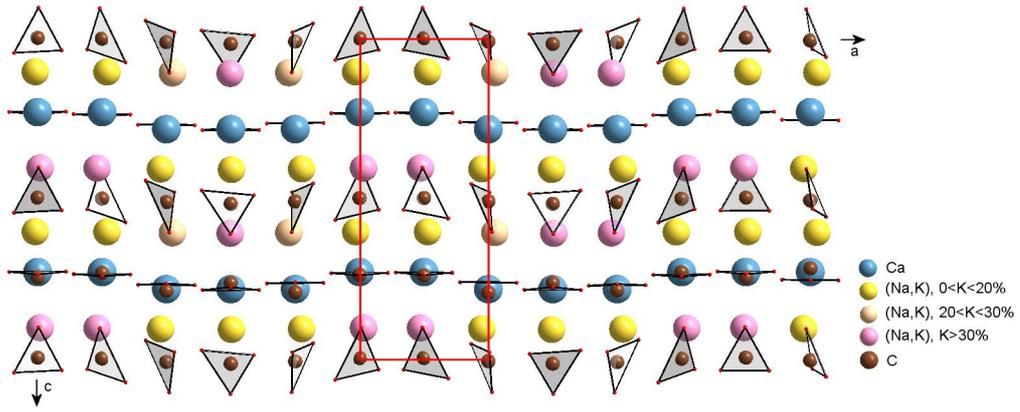

(a)

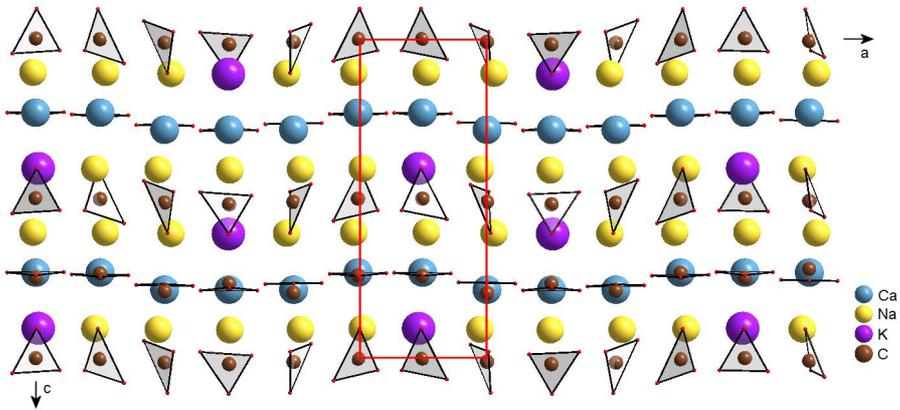

(b)

**Figure 6.** The fragments of the nyerereite structure obtained in models 1 (a) and 2 (b). Unit-cell contours are shown.

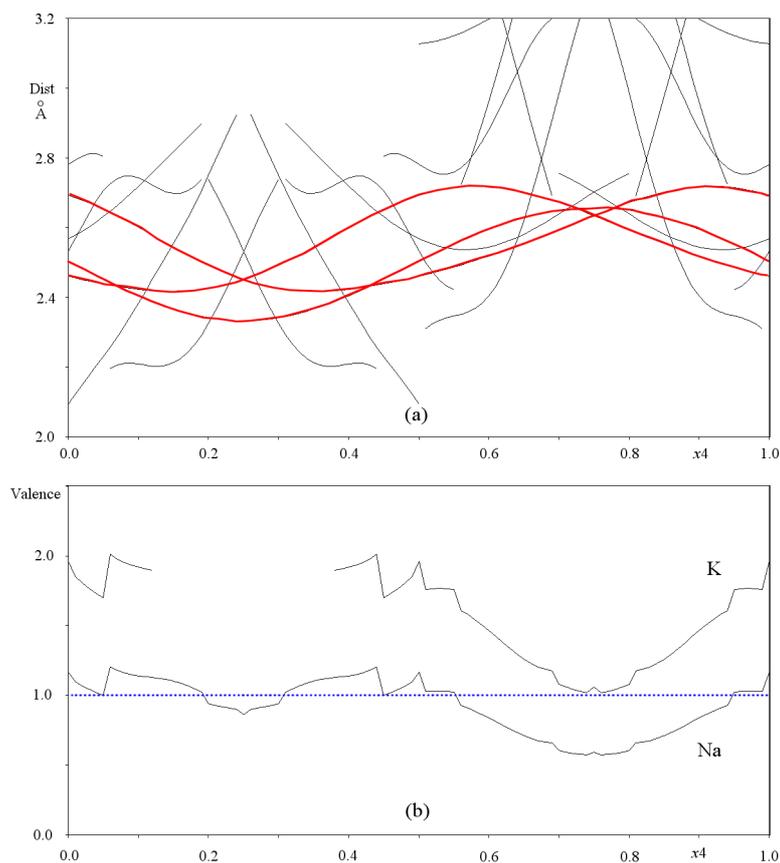

**Figure 7**. (a) (Na,K) – O distances; (b) bond valence sums (BVS) of Na and K atoms. Smoothly changing distances to O1, O2, O2' are shown by red.

Interatomic (Na,K) – O distances (Fig. 7a) and the bond valence sums (BVS) of Na and K atoms (Fig. 7b) were calculated using data of model 1. Distances from (Na, K) sites to O1, O2, O2' change smoothly whereas those to oxygen atoms, located in vertexes of vertical $CO_3$ groups, change quickly and discontinuously. Values of distances near $x4 = 0.75$ are typical for K – O distances what is in a good accord with BVS values in Fig. 7b. Moreover, there is a segment ~0.6 <$x4$ <~0.9 in Fig. 7b, which could well correspond to mixed occupancy of (Na, K) sites, what is an argument in favour of model 1. But at the same time, all distances in Fig. 7a shorten when $x4$-values move away from $x4 = 0.75$, so that only a short $x4$-segment about 0.2 in length centered at 0.75 can be picked out with confidence. It is just the length, which corresponds to model 2 where such a segment is occupied purely by K atoms. One may conclude that model 2 is more probable as compared with model 1.

Interatomic Ca – O distances and the bond valence sums (BVS) of Ca atoms are presented in Fig. 8a, b. The shortest distance is Ca – O3. All the distances, except Ca – O41, change insignificantly on the modulation period.

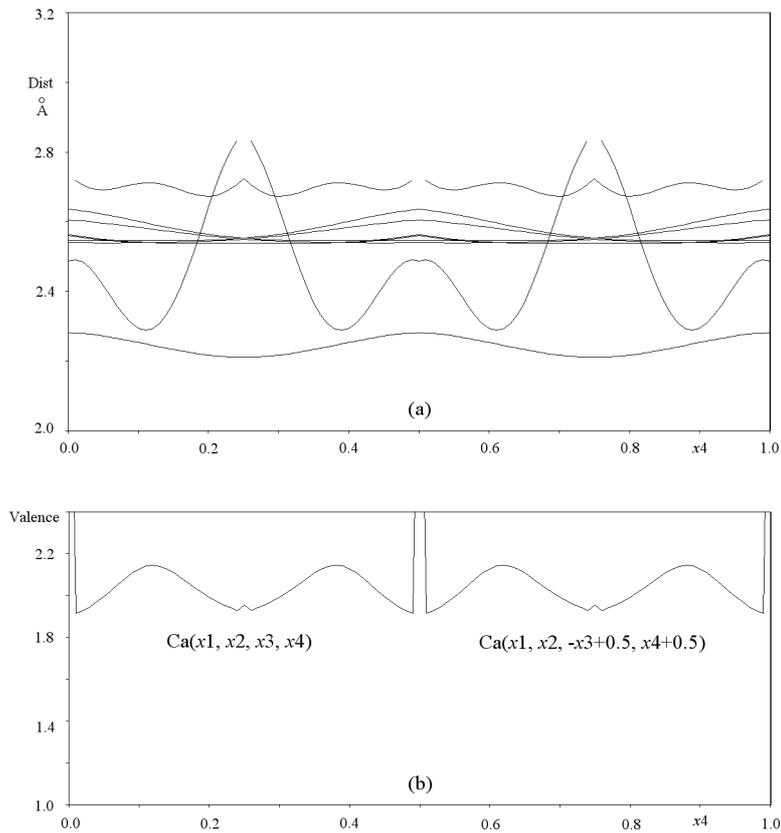

**Figure 8**. (a) Ca – O distances; (b) bond valence sums (BVS) of Ca atoms.

McKie & Frankis (1977) used similar unit cell ($a$ = 5.044, $b$ = 8.809, $c$ = 12.743 Å, Z = 4) but other ($Cmc2_1$) 3D symmetry group to describe average nyerereite structure. This difference is worthy to be discussed. Symmetry operators of the $Cmcm(\alpha00)00s$ and $Cmc2_1$ groups (without those generated by C-centering) are written in two next columns of Table 3. The $Cmc2_1$ group is the non-centrosymmetric subgroup of $Cmcm$ as well as $C2cm(\alpha00)s0s$ (third column) is the non-centrosymmetric subgroup of $Cmcm(\alpha00)00s$.

**Table 3**. (3+1)D versus 3D symmetry of nyerereite

| $Cmcm(\alpha00)00s$ | $Cmc2_1$ | $C2cm(\alpha00)s0s$ |
|---|---|---|
| (1) $x1$; $x2$; $x3$; $x4$ | $x$; $y$; $z$ | $x1$; $x2$; $x3$; $x4$ |
| (2) $-x1$; $-x2$; $x3+0.5$; $-x4+0.5$ | $-x$; $-y$; $z+0.5$ | |
| (3) $-x1$; $x2$; $-x3+0.5$; $-x4$ | | |
| (4) $x1$; $-x2$; $-x3$; $x4+0.5$ | | $x1$; $-x2$; $-x3$; $x4+0.5$ |
| (5) $-x1$, $-x2$; $-x3$; $-x4$ | | |
| (6) $x1$; $x2$; $-x3+0.5$; $x4+0.5$ | | $x1$; $x2$; $-x3+0.5$; $x4+0.5$ |
| (7) $x1$; $-x2$; $x3+0.5$; $x4$ | $x$; $-y$; $z+0.5$ | $x1$; $-x2$; $x3+0.5$; $x4$ |
| (8) $-x1$; $x2$; $x3$; $-x4+0.5$ | $-x$; $y$; $z$ | |

Corresponding extinction rules coincide in the 3D space:

| | |
|---|---|
| $Cmcm(\alpha 00)00s$ & $C2cm(\alpha 00)s0s$ | $Cmc2_1$ |
| $hklm$: $h+k = 2N$ | $hkl$: $h+k=2N$ |
| $hk0m$: $m = 2N$ | |
| $h0lm$: $h = 2N$; $l=2N$ | $h0l$: $h=2N$; $l=2N$ |

There is no good means, however, to describe stepwise $x3$-changes of Ca atoms and horizontal $CO_3$ groups on the modulation period using $Cmc2_1$ as 3D base of a (3+1)D symmetry group, since symmetry-dependent $x3$-variations are limited by $x3$ and $x3 + 0.5$ as opposed to alternate choice $C2cm(\alpha 00)s0s$ based on $C2cm$.

**The relation of $Na_2CO_3$ and $Na_2Ca(CO_3)_2$ crystal structures**

To the best of our knowledge, two carbonates showing incommensurate modulations are known. One of them is well-known γ-$Na_2CO_3$ and other one, nyerereite, is described in the present article. Nyerereite structure may be schematically presented as 'a sum' of sodium carbonate (γ-$Na_2CO_3$) and calcite ($CaCO_3$). All three structures reveal hexagonal motif being projected along their *c*-axes (Fig. 9). The K-free analogue of nyerereite is figured for simplicity.

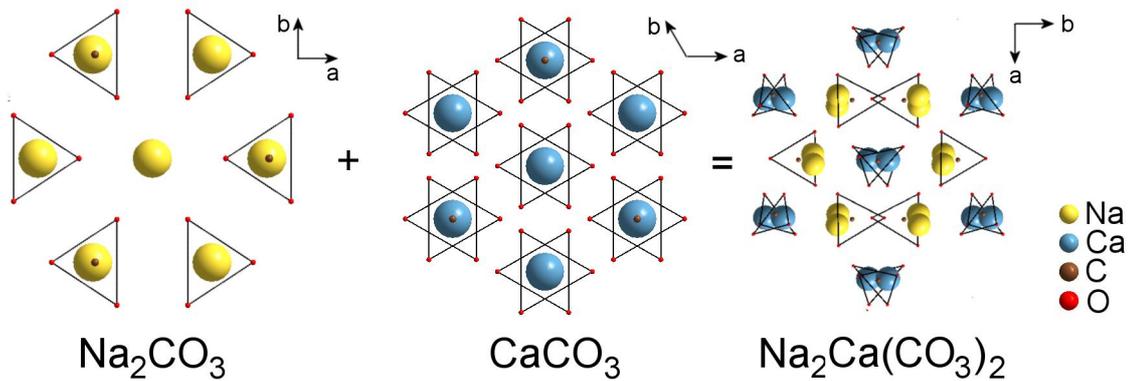

**Figure 9.** A fragment of $Na_2Ca(CO_3)_2$ (Gavryushkin *et al.*, 2016) presented as 'a sum' of γ-$Na_2CO_3$ whose averaged data are taken from (Dusek *et al.*, 2003) and $CaCO_3$ (Chessin, Hamilton & Post, 1965).

The modulation vector **q** = 0.383**a**\* of nyerereite is coaxial with the *a*-axis whereas **q** = 0.18**a**\* + 0.32**c**\* of γ-$Na_2CO_3$ is perpendicular to the *b* axis. Unit-cell periods *a* =5.062, *b* = 8.790 Å of nyerereite are comparable with $b = 5.245$, $a = 8.920$ Å of γ-$Na_2CO_3$. The *c*-period of nyerereite (12.744 Å) is close to the double of *c*-period of γ-$Na_2CO_3$ whose β angle (~101°) is not so far from 90°. A fragment of the nyerereite analogue is oriented so in Fig.9 as to make comparable periods like-directed with those of γ-$Na_2CO_3$. Two modulation vectors are mutually perpendicular in this setting. The largest modulation amplitudes in γ-$Na_2CO_3$ are along its *b*-axis as one can see in Figures 8, 9 from (Dusek *et al.*, 2003). Though considerable modulation amplitudes of (Na,K) sites are determined along comparable *a*-axis of nyerereite, a special feature of this structure is discontinuous modulation of Ca and $CO_3$ groups mainly along the *c*-axis as determined above.

As can be seen from the following analysis, nyerereite and $Na_2CO_3$ structures can be transformed in each other by simple operations. The K-free, commensurate analogue $Na_2Ca(CO_3)_2$ will be considered for simplicity.

The most clearly the structural relations can be seen from the comparison of the cation arrays. As for both compounds high- and low-temperature modifications are isotypic, we illustrate the relations of crystal structures by the example of higher symmetric high-temperature modifications. In the crystal structure of $Na_2CO_3$, there are two topologically different types of Na sites. First of them, Na1, are connected through O atoms with six C atoms and the others, Na2 – with five C atoms (yellow and blue balls in Fig.10(a), respectively). Transformation of Na-C net of $Na_2CO_3$ into (Na,Ca)-C net of $Na_2Ca(CO_3)_2$ can be done in three steps: 1) replacement of 5-coordinated Na2 with Ca, 2) shift of the double layers consisting $NaC_6$ octahedrons, 3) removal of all Ca atoms, whose centers lie on the plane of the shift (Fig.10(c)); this removal is natural because shift splits each 5-coordinated void on two tetrahedral voids, which are too small for such a big atom as Ca. As the shift also splits columns of $NaC_6$ octahedrons, modulated in γ-$Na_2CO_3$, a modulation analogous to that of γ-$Na_2CO_3$ is unlikely both in $Na_2Ca(CO_3)_2$ and nyerereite.

The described shift transforming crystal structure of $Na_2CO_3$ into $Na_2Ca(CO_3)_2$ also gives rise to two non-parallel orientations of $CO_3$ groups. Carbon atoms lying on the plane of the shift change trigonal prismatic coordination (Fig.10(b)) on octahedral one (Fig.10(d)). The $CO_3$ groups inside octahedrons change horizontal orientation into vertical one due to necessity of inclusion (for compensation of bond valences) in the coordination sphere of two Ca atoms, one located above, and another – below the octahedron; see more details in (Gavryushkin *et al.*, 2016). This is noteworthy, that crystal structure of $Na_2Ca(CO_3)_2$ (Fig.10(d)) can be described as the stacking of layers cut from $CaCO_3$ and $Na_2CO_3$ crystal structures. However, layers cut from $CaCO_3$ structure in the structure of $Na_2Ca(CO_3)_2$ have $NaCO_3$ composition and layers from $Na_2CO_3$ structure – $NaCaCO_3$ composition.

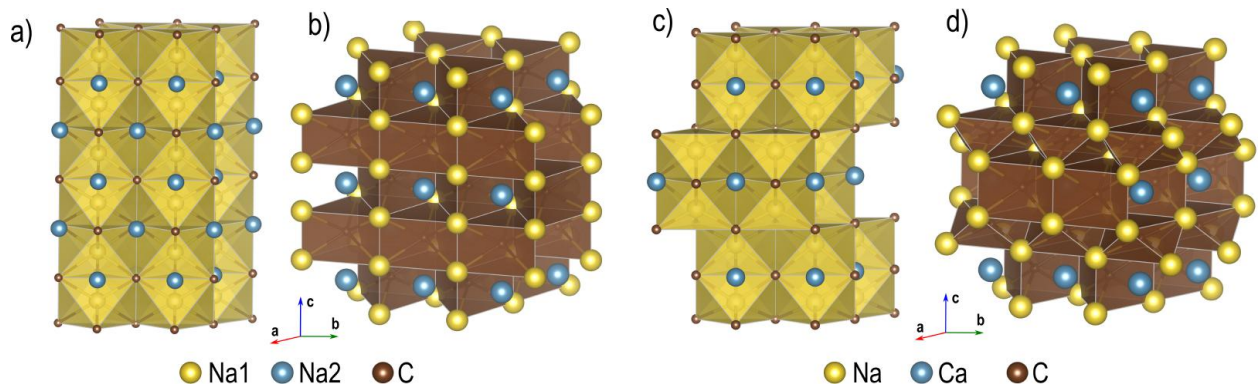

**Figure 10**. Na-C subnet of $Na_2CO_3$ (a,b) and Na-C net of $Na_2Ca(CO_3)_2$ (c,d) with outlined Na-centered (a,c) and C-centered polyhedrons (b,d).

**Conclusions**

The incommensurately modulated twin structure of nyerereite, $Na_{1.64}K_{0.36}Ca(CO_3)_2$, has been solved and refined for the first time using the superspace approach. Main structure difference between natural nyerereite and synthetic $Na_2Ca(CO_3)_2$ studied before is the difference in the modulation type, which is incommensurate for nyerereite but commensurate for its synthetic, K-free counterpart. Both natural and synthetic crystals are twins but differ in relative volumes of their twin components, which are equal in $Na_2Ca(CO_3)_2$ whereas one of the components strongly prevails in nyerereite studied. Varying Na/K occupancy can be described at an acceptable quality level using both complementary harmonics (model 1) and crenels (model 2). At the same time, one may suppose model 2 to be more probable. The supposition is based on the analysis of interatomic (Na,K) – O distances and BVS of Na and K atoms as well. It is worthy of notice that the content of K does not vary in all the mineral samples studied. One may assume that it is not a pure coincidence. A crystal of chemical composition $Na_{1.6}K_{0.4}Ca(CO_3)_2$, whose structure is modulated with **q** = 0.383(1)**a***, can be described by model 2 so that each (or almost each) appropriate (Na, K) site is occupied by the K atom. Finally, the nyerereite structure has been compared with the widely known modulated structure of γ-$Na_2CO_3$.

**Funding**

The research was supported by the Russian Foundation for Basic Research through Grant Nos.

14-05-31051 and 13-0312158 and the Ministry of Education and Science of Russian Federation (Grant Nos. 14.B25.31.0032 and MK- 3766.2015.15). AK and AG were supported by a grant from the Russian Science Foundation (RSF 15-17-30012), and ANZ was supported by grant from St. Petersburg State University (3.38.224.2015).